\documentclass[aps,pre,superscriptaddress,showpacs,amsmath,amssymb,reprint]{revtex4-1}
\usepackage{amsmath}
\usepackage{amsfonts}
\usepackage{amsthm}
\usepackage{graphics}
\usepackage{graphicx}
\usepackage{amssymb}
\usepackage{mathtools}
\usepackage{caption,subcaption}
\usepackage{color}
\usepackage{url}
\usepackage[abs]{overpic}
\usepackage[usenames,dvipsnames]{xcolor}
\usepackage[normalem]{ulem}

\def\kmax{k_{\rm max}}

\newcommand\Rh{{\text{Rh}}}
\newcommand\Rey{{\text{Re}}}

\def\mbfx{\mathbf{x}}
\def\mbfr{\mathbf{r}}
\def\mbfu{\mathbf{u}}

\graphicspath{{./}}

\begin{document}

\title{Equivalence of nonequilibrium ensembles: Two-dimensional turbulence with a dual cascade}

\author{Kannabiran Seshasayanan}
\email{s.kannabiran@gmail.com}
\affiliation{Department of Physics, Indian Institute of Technology Kharagpur, Kharagpur - 721 302, India}%
\author{Karthik Subramaniam Eswaran}
\email{kseswaran@iitkgp.ac.in}
\affiliation{Department of Physics, Indian Institute of Technology Kharagpur, Kharagpur - 721 302, India}%
\author{Maheswar Maji}
\email{mahes.maji88@kgpian.iitkgp.ac.in}
\affiliation{Department of Physics, Indian Institute of Technology Kharagpur, Kharagpur - 721 302, India}%
\author{Sourangshu Ghosh}
\email{sourangshu@iitkgp.ac.in}
\affiliation{Department of Civil Engineering, Indian Institute of Technology Kharagpur, Kharagpur - 721 302, India}%
\author{Vishwanath Shukla}
\email{research.vishwanath@gmail.com}
\affiliation{Department of Physics, Indian Institute of Technology Kharagpur, Kharagpur - 721 302, India}%

\date{\today}
\begin{abstract} 
We examine the conjecture of equivalence of nonequilibrium ensembles for turbulent flows in two-dimensions (2D) in a dual-cascade setup. We construct a formally time-reversible Navier-Stokes equations in 2D by imposing global constraints of energy and enstrophy conservation. A comparative study of the statistical properties of its solutions with those obtained from the standard Navier-Stokes equations clearly show that a formally time-reversible system is able to reproduce the features of a 2D turbulent flow. Statistical quantities based on one- and two-point measurements show an excellent agreement between the two systems, for the inverse- and direct cascade regions. Moreover, we find that the conjecture holds very well for 2D turbulent flows with both conserved energy and enstrophy at finite Reynolds number, which goes beyond the original conjecture for three-dimensional turbulence in the limit of infinite Reynolds number.

\end{abstract}

\maketitle

\section{Introduction}
Equivalence of equilibrium ensembles, such as microcanonical and canonical ensembles at constant energy and at constant temperature, respectively, in the thermodynamic limit (number of constituents $N \to \infty$) is a well known result, usually taught in the first course on statistical mechanics. Can we get such an useful result for systems driven far from equilibrium, especially for turbulence?  Turbulence is a quintessential example of a nonequilibrium dissipative system that involves extremely large number of interacting degrees of freedom. In Ref.~\cite{gallv96PLAequivconj} it was postulated that the statistical properties of turbulence based on the solutions of the Navier-Stokes equation (NSE) and its suitably modified, time-reversal-invariant version, called the reversible Navier-Stokes (RNS) are equivalent in the limit of an infinite Reynolds number (Re). Henceforth ``equivalence conjecture''. The infinite Re limit serves as a thermodynamic limit here.  This conjecture stems from a more general \textit{equivalence of dynamical ensembles} discussed in Ref.~\cite{GallaCohenPRL95} for a sheared fluid in a nonequilibrium steady state.
  
Here we take up the example of a two-dimensional (2D) turbulence and examine the equivalence conjecture more closely, in order to understand its applicability at moderate to large Reynolds numbers. 2D turbulent flows  are often used to understand atmospheric and oceanic flow dynamics that influence the weather and climate systems and whose modeling is of paramount importance currently. It is not just this practical utility of 2D turbulence that is interesting, but it is also one of those turbulence systems that has benefited the most from the ideas emanating from equilibrium statistical mechanics~\cite{kraichnan1967inertial,kraichnan1975remarks}. A time-reversible formulation for 2D turbulence can help in developing coarse-grained LES type models \cite{ShePRL1993}. The inviscid, unforced 2D NSE admits two conserved positive quadratic quantities: the energy and the enstrophy (integrated square of the vorticity). An \textit{ad absurdum} Fj\o rtoft argument illustrates that these two conserved quantities result in a dual cascade behavior in 2D turbulence: the energy injected at scale $\ell_{\rm f}$ cascades towards larger length scales giving rise to  large scale coherent condensate if the friction is not strong while the enstrophy cascades towards smaller length scales.

Are these features reproduced by the formally time-reversal-invariant formulations of the 2D NSE as suggested by the \textit{equivalence conjecture}? In Reference~\cite{gallv96PLAequivconj} a RNS system was built by modifying the (viscous) dissipation term of the NSE, which amounts to constructing an artificial dissipation mechanism that exactly balances the injection statistics of the applied external force so that a prescribed macroscopic observable becomes a constant of motion. The dissipation coefficient constructed in this manner makes the dissipation term in the new governing equation time-reversal invariant. Given the reduced numerical complexity of the 2D NSE, the validity of the conjecture has been examined in the past, albeit for few to moderate number of degrees of freedom. The consequences of the chaotic hypothesis were tested for 2D flows~\cite{rondoniRevdisp2Dturb99,gallv2004lyapunov}, wherein fluctuations of global quadratic quantities were examined in statistically stationary states while keeping one global quadratic quantity fixed. Comparison of the Lyapunov spectra showed that these match with the ones computed using NSE, thereby suggesting that the conjecture holds. However, the conjecture must be examined in numerical simulations of 2D turbulent flows where both the cascades are well resolved and higher order statistics are studied in detail.% 

Studies in three-dimensions (3D) based on reduced models and direct numerical simulations of RNS systems indicate a growing support for the conjecture~\cite{ShePRL1993,biferaleTimerev98,depietro2018shell,RNS3dVS19,RNSChibbaro21}. A time-reversible shell model of turbulence~\cite{biferaleTimerev98}, obtained by imposing a global constraint of energy conservation, was used to explore a part of the RNS parameter space by varying the forcing strength; the system underwent a smooth transition from an equilibrium state to a nonequilibrium stationary state which exhibit an energy cascade from large to small length scales. This important work was further extended in Ref.~\cite{RNS3dVS19} wherein different statistical regimes of the RNS system (for a constant energy constraint) were systematically examined in DNSs at a modest resolution based on quantities that derive from one-point measurements. It suggested that the RNS systems could perhaps provide a framework, which is capable of yielding genuine turbulent statistics out of a time-reversible dynamics. Moreover, DNS results, aided by the Leith model and a heuristic mean-field Landau free energy argument, conclusively demonstrated the existence of continuous second-order phase transition from warm states (partially thermalized solutions) to  hydrodynamic states with compact energy support in k-space (and insensitive to the cut-off scale), the (normalized) enstrophy or equivalently reversible-dissipation coefficient plays the role of an order parameter. 

Recent works have also explored the consequences of global constraints other than the total energy, where it has been suggested that the choice of a global  macroscopic observable is an important factor in determining the relevance of the associated RNS system for comparison with the NSE~\cite{biferale2018equivconj}. Moreover, the time-reversible shell model was also used to study the time irreversibility that is associated with the nonlinear energy transfers (energy cascade) in 3D, given that the RNS system avoids the explicit time-reversal symmetry breaking by construction~\cite{depietro2018shell}. Recently, a model obtained by imposing the constraint that turbulent enstrophy is conserved has been analyzed in 3D turbulence where the RNS and NSE systems were shown to be similar~\cite{RNSChibbaro21}.

The primary objective of the present work is to test the validity of the \textit{equivalence conjecture} for 2D turbulence. We perform well resolved DNSs of the 2D NSE and RNS in a dual cascade setup and compare statistical properties of their solutions using one- and two-points measurements. Similar to the 3D case~\cite{RNS3dVS19}, we find that 2D RNS reproduces very well the global observables and is able to capture the essential features of the two cascade processes such as energy spectra and fluxes, which match with their NSE counterparts. The third-order longitudinal velocity structure function obtained using the RNS system exhibits scaling behaviour similar to what is seen for the 2D NSE. 

The remainder of this paper is organized as follows. In Sec.~\ref{sec:theo2DRNS} we give an overview of the 2D RNS system as appropriate for this work. Section~\ref{Sec:numsetup} provides a summary of the numerical methods and parameters that are used in this study, along with a brief account of various statistical quantities that we use. Section~\ref{sec:resultsdiscuss} contains results of our numerical simulations and their discussion, while in Sec.~\ref{sec:conclusions} we give the conclusions drawn from results and discuss the significance of our work.

\section{Time-reversible 2D Navier-Stokes equations}
\label{sec:theo2DRNS}

The two-dimensional Navier-Stokes equation is written in terms of the velocity field ${\bf u}$ as
\begin{equation}\label{eq:2dnseu}
\frac{\partial \mbfu}{\partial t} + \mbfu\cdot\nabla\mbfu = -\frac{1}{\rho}\nabla p + \nu \nabla^{2n}\mbfu - \alpha \mbfu + \mathbf{f}_{u},
\end{equation}
where $p$ is the pressure field, $\nu$ is the kinematic viscosity for $n=1$, $\alpha$ is the linear friction damping, and $\mathbf{f}_u$ is the forcing term. The incompressibility condition is ensured by requiring $\nabla\cdot \mbfu =0$. Also, expressed in terms of its components the velocity field is $\mbfu(\mbfx) = (u_x(\mbfx),u_y(\mbfx))$. By introducing the stream function $\psi$, the velocity field can be written as $\mbfu \equiv (\partial_y \psi,-\partial_x \psi)$. Thereby, the above NSE can be expressed in stream function and vorticity formulation as
\begin{equation}\label{eq:2dnsew}
\frac{\partial \omega}{\partial t} + J(\omega,\psi) = \nu \nabla^{2n}\omega - \alpha \omega + f_{\omega},
\end{equation}  
where $\omega= \widehat{\bf e}_z \cdot \left( \nabla \times \mbfu \right) = -\nabla^2 \psi$ is the vorticity field, $J(\omega,\psi)=\mbfu\cdot\nabla\omega=\partial_x\omega\partial_y\psi-\partial_y\omega\partial_x\psi$, and $f_{\omega} {\bf e}_z=\nabla\times \mathbf{f}_u$.

In the presence of viscous dissipation and large scale friction the resulting macroscopic dynamics described by the 2D NSE is irreversible. The NS Eq.~\eqref{eq:2dnseu} is not invariant under the transformation
\begin{equation}
	\mathcal{T}_u: t \to -t; \quad \mbfu \to -\mbfu;
\end{equation}
equivalently Eq.~\eqref{eq:2dnsew} (vorticity formulation) is not invariant under the transformation $\mathcal{T}_{\omega}: t \to -t; \omega \to - \omega$. However, following the equivalence conjecture, coefficients of the dissipative terms can be modified to yield a formally time-reversible governing equation, which is invariant under the transformation $\mathcal{T}_u$ (and equivalently $\mathcal{T}_{\omega}$). 

For the 2D system under consideration, if we decide to impose the global conservation of the total energy and enstrophy, we can construct a desired RNS system with dissipation coefficients $\alpha_r$ and $\nu_r$, that are chosen in order to conserve both the energy and enstrophy. Equations~\eqref{eq:2dnseu} and \eqref{eq:2dnsew} can be used to derive the following balance equations:
\begin{subequations}
	\begin{align}
	\frac{\partial \langle \mbfu^2 \rangle}{\partial t} &= -\nu \langle \omega^2 \rangle - \alpha \langle \mbfu^2 \rangle + \langle \mbfu\cdot \mathbf{f}_u \rangle; \label{eq:energy_bal}\\
	\frac{\partial \langle \omega^2 \rangle}{\partial t} &= - \nu \langle | \nabla \times \left( \omega \hat{e}_z \right) |^2\rangle - \alpha \langle \omega^2 \rangle + \langle \omega f_{\omega} \rangle; \label{eq:enstro_bal}
	\end{align}
\end{subequations}
wherein $\langle \cdot \rangle$ denotes spatial averaging.

Let $E \equiv \langle \mbfu^2\rangle$ denote the energy, $\Omega \equiv \langle \omega^2 \rangle$ the enstrophy, and $P \equiv \langle | \nabla \times \left( \omega \hat{e}_z \right) |^2\rangle$ the palinstrophy. Also, the energy and enstrophy injection rates are given by $\varepsilon_I = \langle \mbfu\cdot \mathbf{f}_u \rangle$ and $\eta_I= \langle \omega f_{\omega} \rangle$. For the RNS system conserving both the total energy and the enstrophy, we have the relations $dE/dt = d\Omega/dt = 0$.
Substituting the relations into Eqs.~\eqref{eq:energy_bal} and ~\eqref{eq:enstro_bal} gives $\varepsilon_I = \alpha E + \nu \Omega$ and $\eta_I = \alpha \Omega + \nu P$. Therefore, the coefficients $\alpha_r$ and $\nu_r$ for the 2D RNS system, conserving both the energy and enstrophy, are given by
\begin{subequations}\label{eq:tcoeffsrns}
	\begin{align}
	\alpha_r [\mbfu] &= \frac{\varepsilon_{I} P - \eta_I \Omega}{E P - \Omega^2}, \label{eq:alphar} \\
	\nu_r [\mbfu] &= \frac{\varepsilon_I \Omega - \eta_I E}{\Omega^2 - P E}. \label{eq:nur}
	\end{align}
\end{subequations}
Observe that under the transformation $\mathcal{T}_u$ we get $\varepsilon_I \to -\varepsilon_I$ and under the transformation $\mathcal{T}_{\omega}$ we get $\eta_I \to -\eta_{I}$. This change of sign under these transformations makes the dissipative terms $\alpha_r \mbfu$ and $\nu_r \nabla^2 \mbfu$ invariant under $\mathcal{T}_u$. Also note that the \textit{reversible friction coefficient} $\alpha_r$ and \textit{reversible viscosity coefficient} $\nu_r$ are functionals of $\mbfu$, thus depend on the state of the system. The RNS system is obtained by replacing the constant in time values of $\alpha$ and $\nu$ in the NSE with the state dependent $\alpha_r$ and $\nu_r$, respectively:
\begin{equation}\label{eq:2drnseu}
	\frac{\partial \mbfu}{\partial t} + \mbfu\cdot\nabla\mbfu = -\frac{1}{\rho}\nabla p + \nu_r \nabla^{2}\mbfu - \alpha_r \mbfu + \mathbf{f}_{u},
\end{equation}
incompressibility is enforced by demanding $\nabla \cdot \mbfu=0$. In the turbulent/chaotic regime, state dependent coefficients $\alpha_r, \nu_r$ are functions of time since the velocity field is a time-dependent field.

\section{Numerical Setup}
\label{Sec:numsetup}

We compare the statistical properties of the solutions of the NSE~\eqref{eq:2dnseu} and the RNS Eq.~\eqref{eq:2drnseu} in a turbulent flow at statistical steady state. In order to do this comparison, we perform direct numerical simulations of the the governing equations in the vorticity formulation (see e.g., Eq.~\eqref{eq:2dnsew}) using a highly accurate pseudo-spectral method with periodic boundary conditions~\cite{gomez2005parallel}. We consider a square simulation domain of dimensions $2\pi L \times 2 \pi L$ where $L$ is a length scale of the system. We use $N^2_c$ collocation points and the $2/3$-dealiasing rule; the maximum wavenumber is $k_{max}=N_c/3$. For time integration we use a standard Runge-Kutta RK443 scheme~\cite{ascher1997implicit}. For the RNS simulations, the time step is kept small in order to minimize the energy- and enstrophy-conservation errors during the numerical time integration, see Tab.~\ref{table:runs_values} for an estimate of these errors.

Steady state turbulent flow-field is maintained by forcing the vorticity field with a spatially periodic forcing which has been studied elsewhere \cite{seshasayanan2014edge}, given by
\begin{equation}
	f_{\omega}(\mbfx) = 2 \tilde{k}_f f_{0} \sin(\tilde{k}_f \,x)\, \sin(\tilde{k}_f \,y),
\end{equation}
where $f_0$ and $\tilde{k}_f$ are, respectively, the forcing amplitude and  wavenumber. We write $k_f \coloneqq \sqrt{2} \tilde{k}_f$ as norm of the forcing wave vector $(\tilde{k}_f,\tilde{k}_f)$. The NSE runs are started using the initial condition
\begin{equation}
	\omega(\mbfx) = \omega_0\,\sum^{18}_{k_i=14}2 k^2_i\sin(k_i\, x + \theta_x ) \, \sin( k_i\, x + \theta_y ),
\end{equation}
where $\theta_x$ and $ \theta_y $ are random phases. Here $\omega_0$ is a constant which is used to fix the initial energy. 

We carry out the DNSs of the NSE and RNS systems on collocation points up to $N^2_c=4096^2$. The forcing wavenumber is fixed at $k_f L=32\sqrt{2}$. Therefore, we work in the dual cascade setup, where we allow for almost a decade of wavenumber range on either side of the forcing wavenumber.

In this work our goal is to examine the statistical properties of the RNS system and provide a comparative assessment with respect to the NSE. Therefore, we use the following protocol for performing the DNSs. We allow the standard NSE system to attain a statistically steady state at a fixed viscosity, friction and forcing strength. Then the steady state solution of the NSE at a given instant is fed to the RNS system, wherein we keep the forcing strength same and allow the friction and viscous dissipation coefficients to fluctuate in time, as dictated by the dynamical evolution. In order to better compare the two systems we non-dimensionalize the variables in both the NSE and RNS systems using the length scale $L$, velocity scale $U$ and a time scale $L/U$. Here $U$ is the rms velocity taken from the NSE system and is defined as $U = \left\langle |{\bf u}|^2 \right\rangle_{ {\bf x}, t}^{1/2}$; $\left\langle  \right\rangle_{ {\bf x}, t}$ denotes both time and spatial averaging. We are interested in three non-dimensional parameters of the system that can be constructed from Equation~\eqref{eq:2dnsew}. They are the Reynolds number ($\Rey$), large scale Reynolds number ($\Rh$) and the forcing wave number $k_f$. The two Reynolds numbers are defined as,
\begin{align}
\Rey = \frac{U L}{\nu}, \quad\quad \Rh = \frac{U}{L \alpha}. 
\end{align}
We define $\nu_{_\text{NSE}} = 1/\Rey$ and $\alpha_{_\text{NSE}} = 1/\Rh$ which are constant values that will be later compared with the non-dimensional $\nu_{_\text{RNS}} (t) = {\nu_r (t)}/{U L}$ and $\alpha_{_\text{RNS}} (t) = {\alpha_r (t)}/{U L}$. We retain the same notation for all other quantities after non-dimensionalization for the purpose of simplicity. Table~\ref{table:runs_values} shows the numerical resolutions that were used to simulate the 2D turbulent flow along with the parameters $Re$ and $Rh$ that were explored in this study.

\begin{table}
	\begin{center}
		\begin{tabular}{|c| c | c | c | c |} 
			\hline
			$N^2_c$& Re & Rh & $\frac{\Delta E}{E}$ & $\frac{\Delta \Omega}{\Omega}$\\ [0.5ex] 
			\hline
			$256^2$ & $1.7 \times 10^2$ &  $2.3 \times 10^4$ & $2.8 \times 10^{-5}$ & $4.8 \times 10^{-5}$\\ 
			\hline
			$256^2$ & $4.9 \times 10^2$ & $2.2 \times 10^4$ & $4.1 \times 10^{-4}$ & $1.2 \times 10^{-3}$\\
			\hline
			$256^2$ & $2.3 \times 10^3$ & $2.1 \times 10^4$ & $3.4 \times 10^{-4}$ & $1.2 \times 10^{-2}$\\
			\hline
			$512^2$ & $7.2 \times 10^3$ & $2.2 \times 10^4$ & $1.7 \times 10^{-5}$ & $3.4 \times 10^{-3}$\\ [1ex] 
			\hline
			$512^2$ & $1.7 \times 10^4$ & $2.3 \times 10^4$  & $4.9 \times 10^{-7}$ & $1.2 \times 10^{-4}$\\ [1ex] 
			\hline
			$512^2$ & $2.3 \times 10^4$ & $2.2 \times 10^4$  & $9.6 \times 10^{-7}$ & $2.9 \times 10^{-4}$\\ [1ex] 
			\hline
			$1024^2$ & $3.7 \times 10^4$ & $2.2 \times 10^4$  & $8.6 \times 10^{-7}$ & $7.9 \times 10^{-4}$\\ [1ex] 
			\hline
			$1024^2$ & $5.0 \times 10^4$ & $2.2 \times 10^4$ & $4.7 \times 10^{-8}$ & $4.8 \times 10^{-4}$\\ [1ex] 
			\hline
			$4096^2$ & $7.8 \times 10^4$ & $2.2 \times 10^4$ & $9.0 \times 10^{-8}$ & $2.9 \times 10^{-4}$\\ [1ex] 
			\hline
		\end{tabular}
	\end{center}
	\caption{Table of runs showing the number of collocation points, $N^2_c$,  the Reynolds number, $\Rey$ and  the large scale Reynolds number, $\Rh$. $\Delta E/E$ and $\Delta \Omega/\Omega$ provide an estimate of the non-conservation of the imposed global constraints of conserved total energy $E$ and enstrophy $\Omega$ for the RNS system.}
	\label{table:runs_values}
\end{table}

Hereunder, we define some of the statistical quantities that will be used to describe the state of the system. The isotropic energy spectrum is
\begin{equation}
	E(k,t) \coloneqq \sum_{k - \frac{1}{2} < |k'| \leq k + \frac{1}{2}} |\hat{\mbfu}(\mathbf{k'},t)|^2.
\end{equation}
where $\widehat{\bf u}$ denotes the two-dimensional Fourier transform of the velocity field. The total energy and enstrophy are given by,
\begin{subequations}
\begin{align}
E(t) = \int \int {\rm dx \, dy} |{\mbfu} ({\bf x}, t)|^2  = \sum_{|\mathbf{k}|=0}^{\kmax} |\hat{\mbfu}(\mathbf{k},t)|^2, \\
\Omega (t) = \int \int{\rm dx \, dy}  |\omega ({\bf x}, t)|^2  = \sum_{|\mathbf{k}|=0}^{\kmax} k^2 |\hat{\mbfu}(\mathbf{k},t)|^2.
\end{align}
\end{subequations}

The fluxes of energy and enstrophy are given by $\Pi_E(k) = \sum^{k}_{0} \mathcal{T}(\mathbf{k},t)$ and $\Pi_Z(k) = \sum^{k}_{0} \mathcal{Z}(\mathbf{k},t)$, respectively, wherein the nonlinear transfer functions are computed as %(\ADDP{check the expressions})
\begin{subequations}
	\begin{align}
		\mathcal{T}(k,t) &=  \left\langle {\bf u}_{k} \cdot \left( \left( {\bf u} \cdot \nabla \right) {\bf u} \right) \right\rangle \\
		\mathcal{Z}(k,t) &= \left\langle {\omega}_{k} \left( \left( {\bf u} \cdot \nabla \right) {\omega} \right) \right\rangle,
	\end{align}
\end{subequations}
where ${\bf u}_{k}$ denotes the velocity field constructed from modes with wavenumber $k$ and $\omega_{k}$ denotes the vorticity field constructed similarly.   

In the present work we also examine some of the two-point statistical properties of the 2D turbulent flow. The velocity increments between two points separated by a distance $\mbfr$ are given by
\begin{subequations}
\begin{align}
\delta {\bf v}(\mbfr, t) &= \mathbf{u}(\mbfx+\mbfr, t) - \mathbf{u}(\mbfx, t),\\
\delta v_{\parallel}(\mbfr, t) &= \delta {\bf v} (\mbfr, t)\cdot \hat{\mbfr},
\end{align}
\end{subequations}
where $\delta v_{\parallel}$ is the longitudinal velocity increment and $\hat{\mbfr}=\mbfr/|\mbfr|$. Due to the isotropic nature of the turbulent flow we expect that $\delta v_{\parallel} (\mbfr, t)$ to be a function of $r$, independent of $\hat{\bf r}$.  We will be concentrating only on the longitudinal velocity increment so we will denote $\delta v_{\parallel}$ simply as $\delta v$. The increments can be used to compute the $p$th-order longitudinal structure functions
\begin{equation}
S^{(p)}_L(r) = \langle [\delta v(r, t)]^p \rangle_{ {\bf x}, t}.
\end{equation} 

\section{Results and discussion}
\label{sec:resultsdiscuss}

\begin{figure*}
	\includegraphics[scale=0.35]{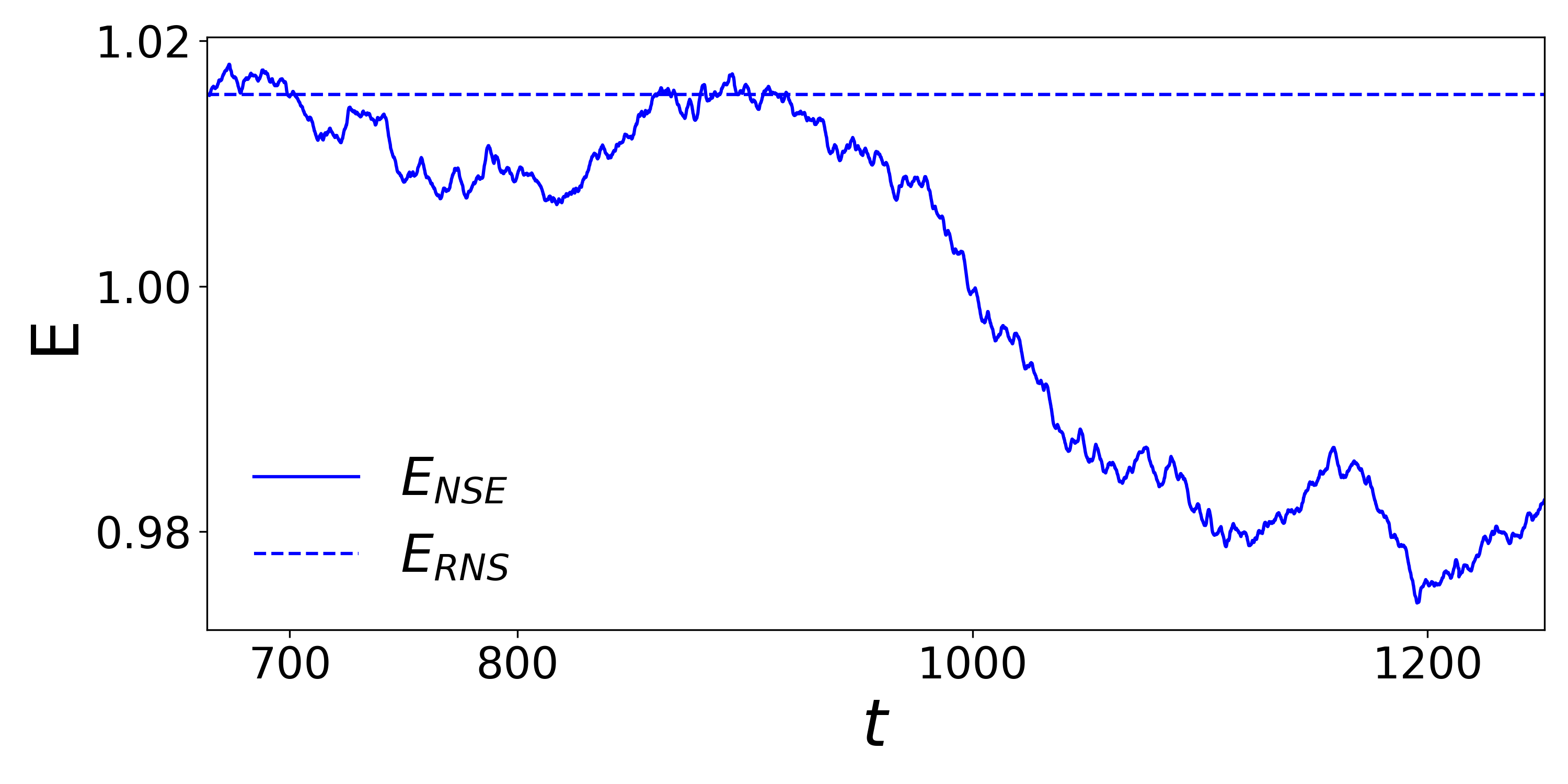}
	\put(-40,105){\Large (a)}
	\includegraphics[scale=0.35]{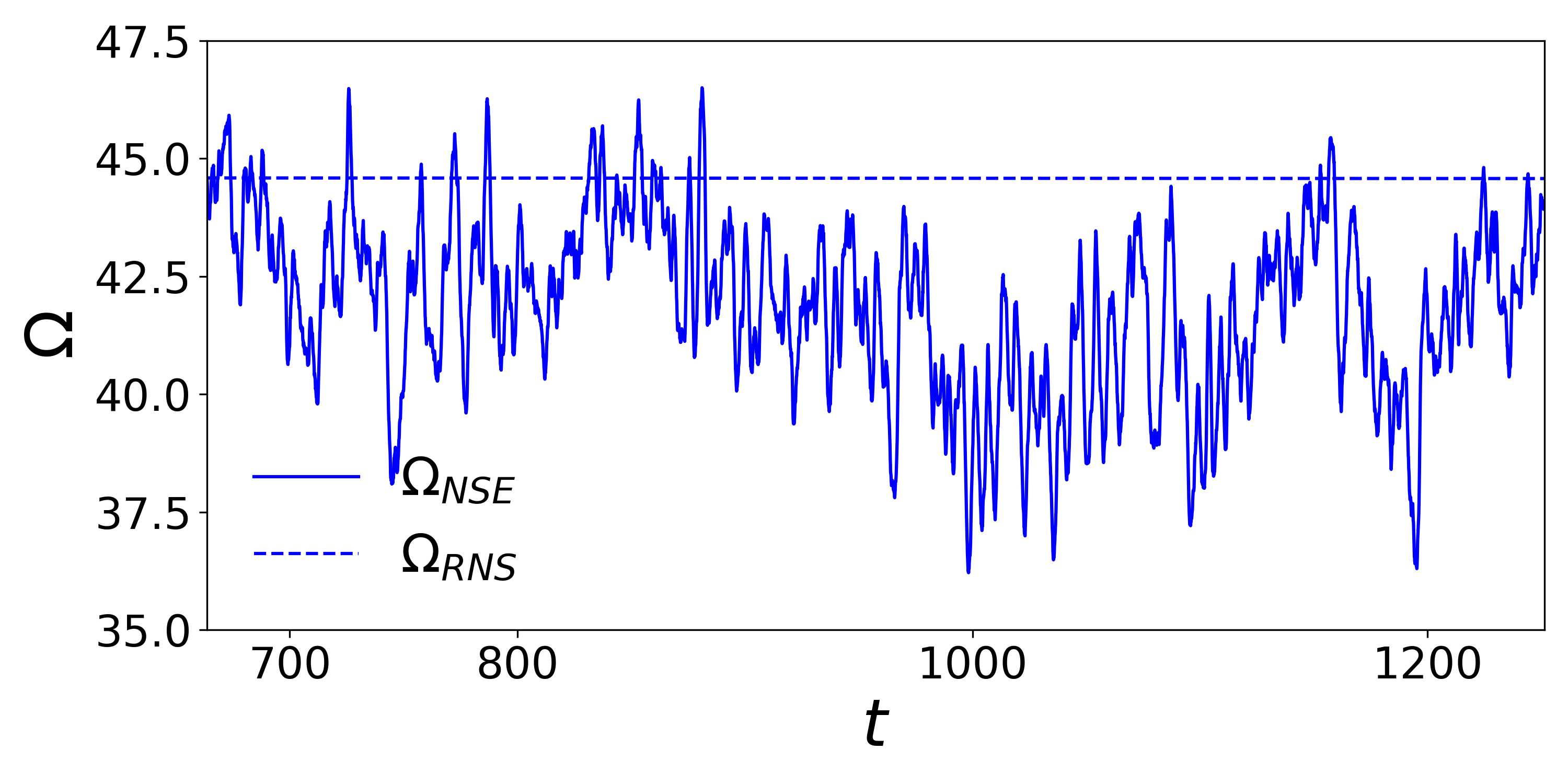}
	\put(-40,105){\Large (b)}\\
	\includegraphics[scale=0.35]{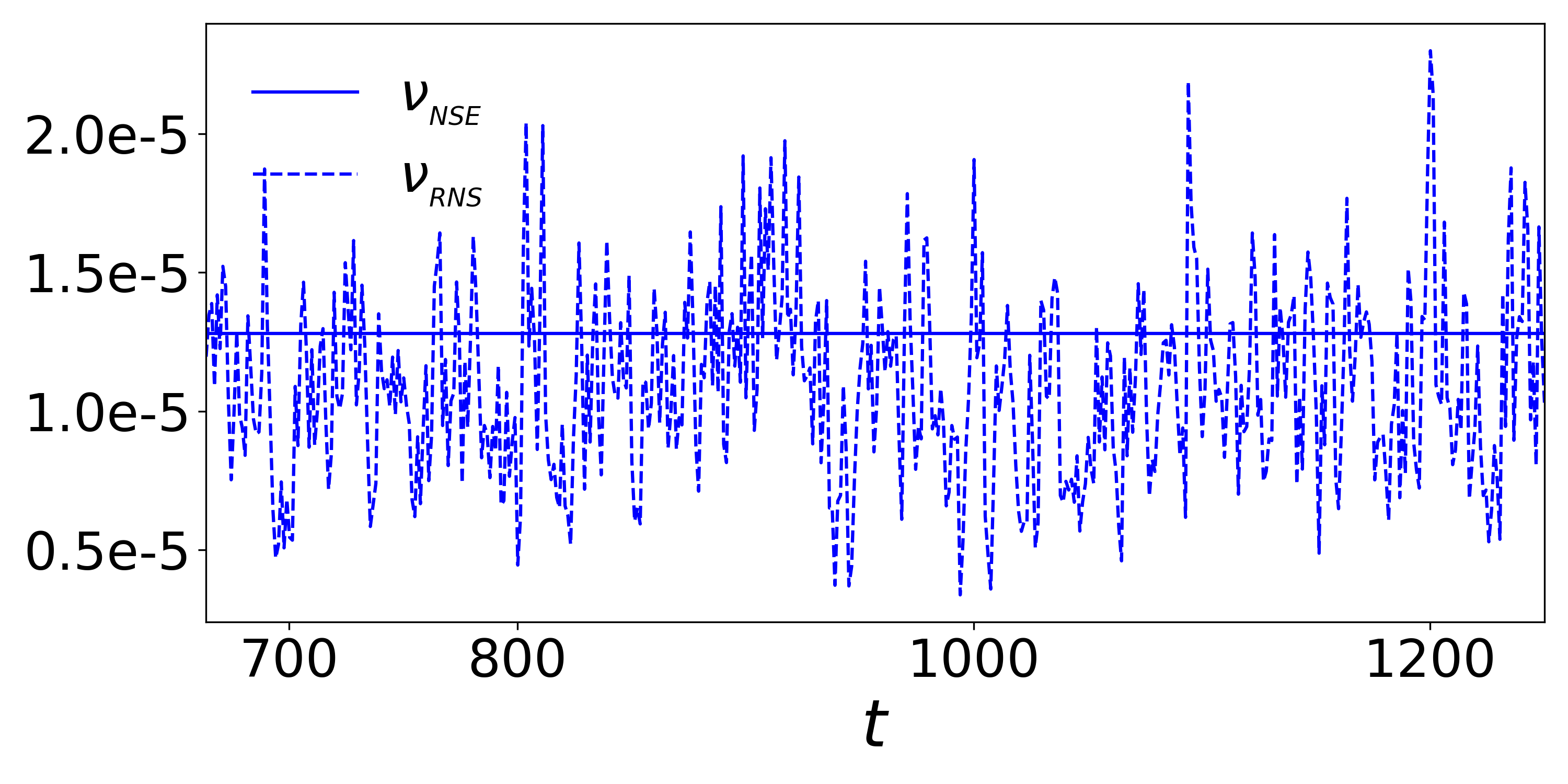}
	\put(-40,105){\Large (c)}
	\includegraphics[scale=0.35]{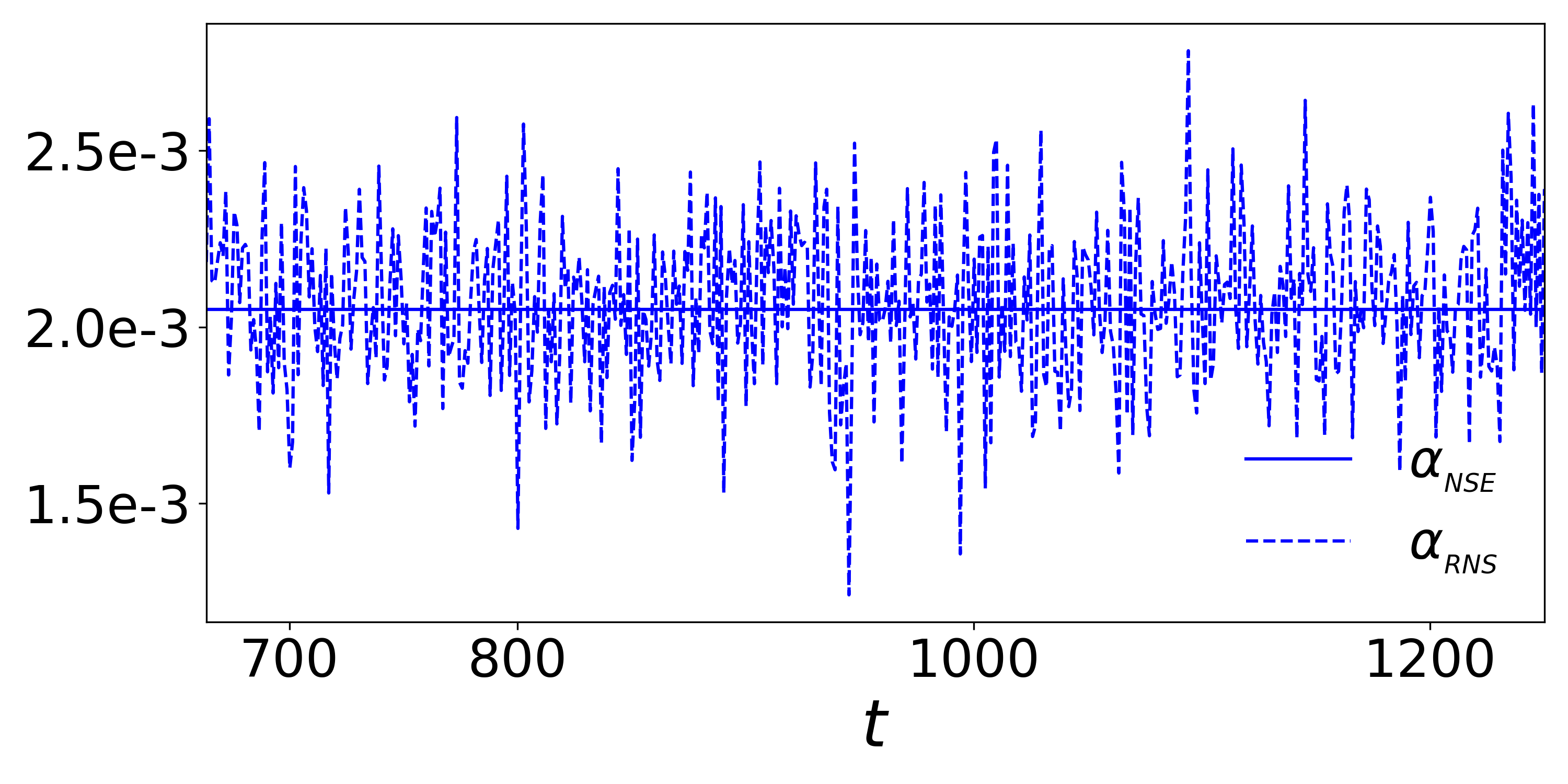}
	\put(-40,105){\Large (d)}
	\caption{Plots of time series from NSE (solid lines) and RNS (dashed lines) systems showing: (a) Energy $E$; (b) enstrophy $\Omega$; (c) viscosity $\nu$; (d) large-scale friction $\alpha$. The runs correspond to $\Rey \approx 7.8 \times 10^4$ and $\Rh \approx 2.2 \times 10^4$ with $N^2_c=4096^2$ collocation points. }
	\label{fig:timeseries}
\end{figure*}

\begin{figure*}
	%\resizebox{\linewidth}{!}{%
	\includegraphics[scale=0.35]{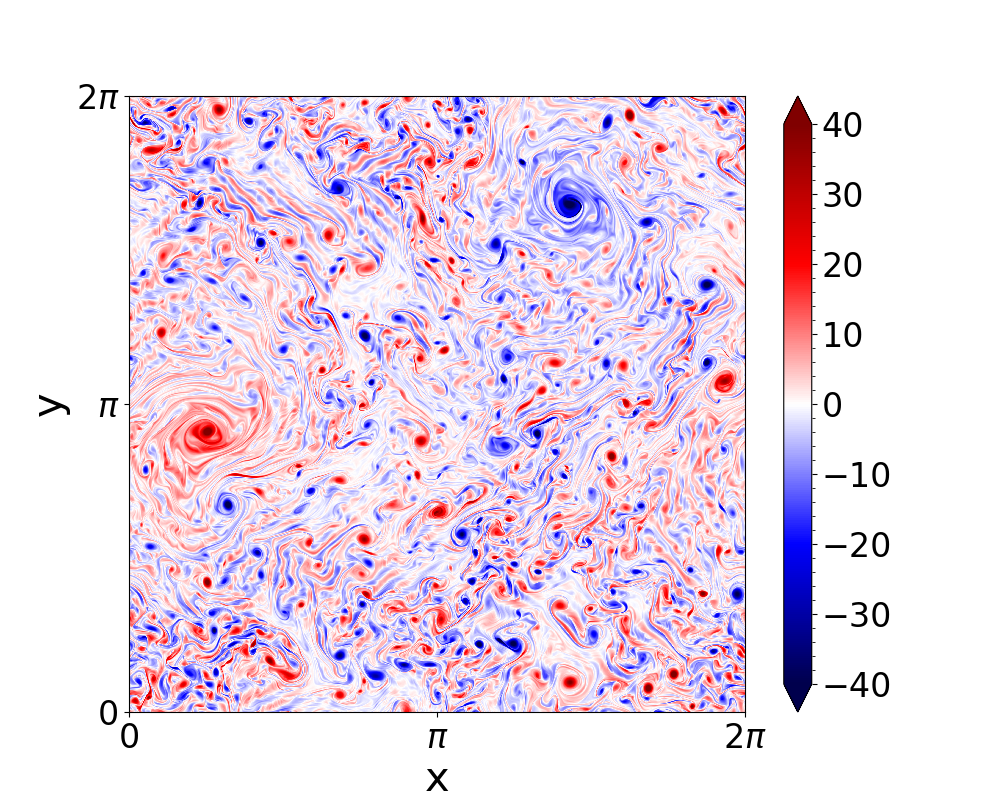}
	\put(-255,170){\Large (a)}
	\includegraphics[scale=0.35]{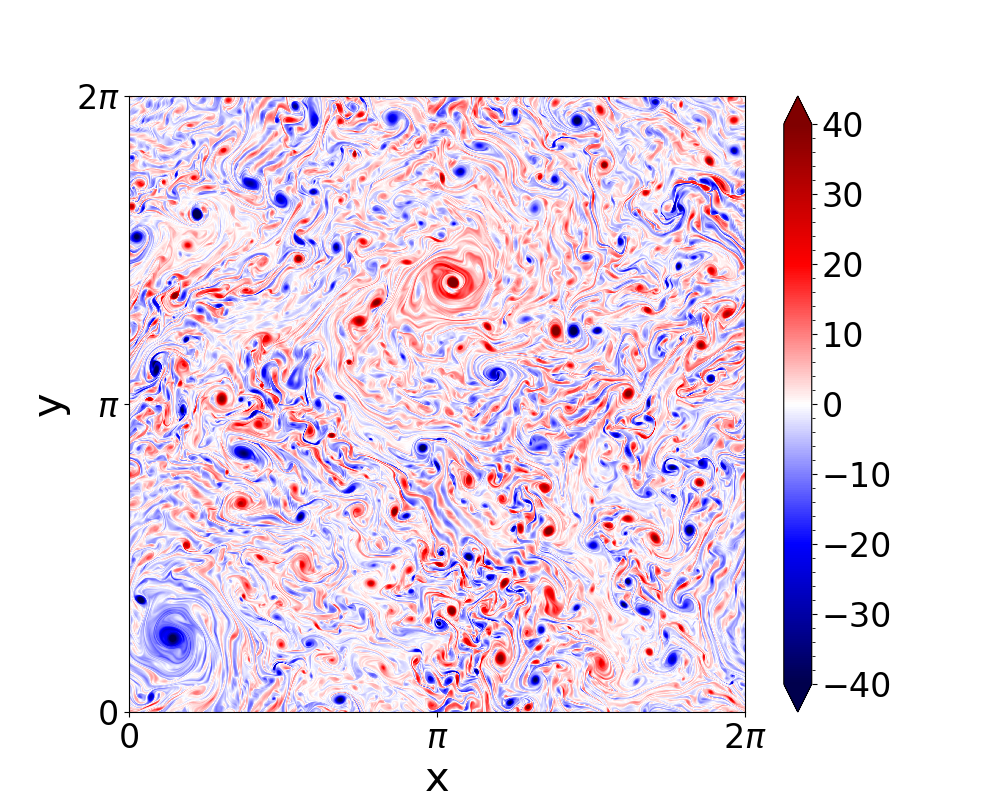}
	\put(-255,170){\Large (b)}
	%}
	\caption{Pseudocolor plots of vorticity for: (a) NSE and (b) RNS simulations at $\Rey \approx 7.8 \times 10^4$ and $\Rh \approx 2.2 \times 10^4$ with $N^2_c=4096^2$ collocation points.}
	\label{fig:vortsnap}
\end{figure*}

The RNS system has been obtained by imposing the global constraints of constant energy $E$ and enstrophy $\Omega$. Therefore, it is imperative to compare the energy $E$ and enstrophy $\Omega$ with the NSE counterparts. Figure~\ref{fig:timeseries} (a) and (b) show the time series of $E$ and $\Omega$ for both the NSE and RNS systems. We see that the energy and enstrophy conservation relations are well satisfied by the RNS system while the $E, \Omega$ of the NSE fluctuates around the $E, \Omega$ values of the RNS respectively. Thus, we can say that $E_{RNS} \approx \langle E_{NSE} \rangle_t$ and $\Omega_{RNS} \approx \langle \Omega_{NSE} \rangle_t$. Similarly, Fig.~\ref{fig:timeseries} (c) and (d) show the time series of the state dependent diffusion coefficients  $\nu_{_\text{RNS}}$ and $\alpha_{_\text{RNS}}$ (see Eq.~\ref{eq:tcoeffsrns}), respectively. Clearly, as discussed before, these quantities fluctuate in time, but their time-average is equal to the constant values that are used in the NSE, $\langle \alpha_{_\text{RNS}} \rangle_t \approx \alpha_{_\text{NSE}} (= 1/\Rey)$ and $\langle \nu_{_\text{RNS}} \rangle_t \approx \nu_{_\text{NSE}} ( = 1/\Rh)$. 
Moreover, for the simulations at large Reynolds numbers of $\Rey \approx 7.8 \times 10^4, \Rh \approx 2.2 \times 10^4$ , with well resolved dissipation ranges, values of $\alpha_{_\text{RNS}}$ and $\nu_{_\text{RNS}}$ are never negative for the entire duration of simulation that we explored. 

Figures~\ref{fig:vortsnap} (a) and (b) show the pseudocolor plots of the vorticity field $\omega$ for the NSE and RNS systems, respectively, for $\Rey \approx 7.8 \times 10^4, \Rh \approx 2.2 \times 10^4$, and forced at an intermediate length scale $k_f = 32 \sqrt{2}$. Both the vorticity fields exhibit similar features, including a dominant pair of a positive and negative vortex, which is usually associated with an inverse cascade of energy if the large-scale friction is not large.

\begin{figure*}
	\resizebox{\linewidth}{!}{%
		\includegraphics[scale=0.32]{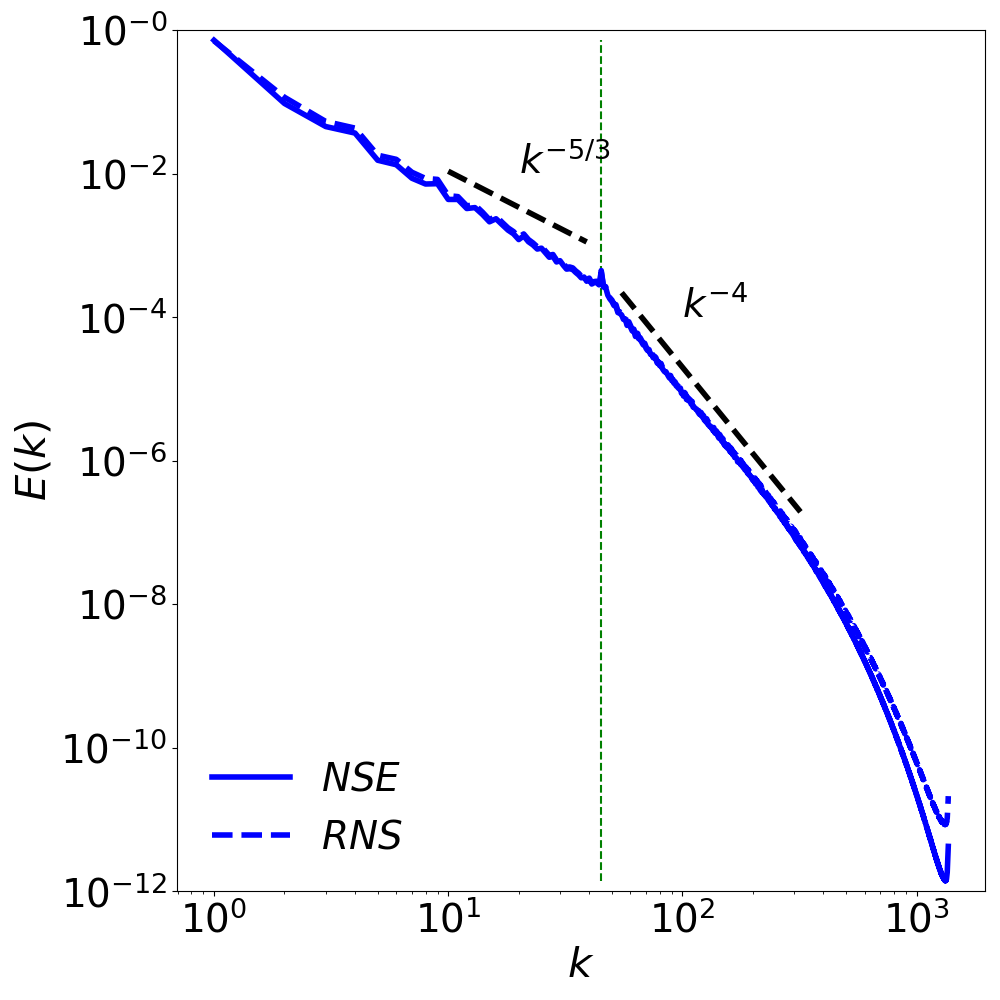}
		\put(-30,190){\Large (a)}
		\includegraphics[scale=0.32]{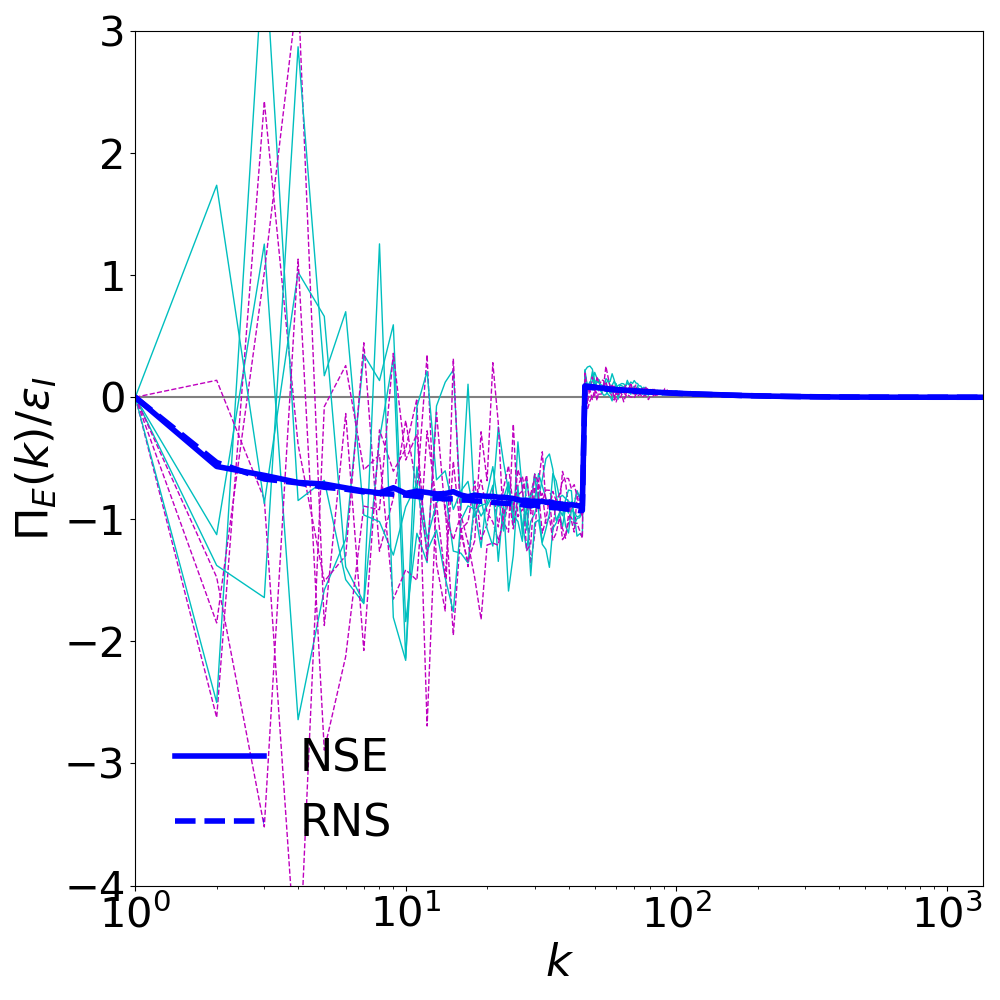}
		\put(-30,190){\Large (b)}
		\includegraphics[scale=0.32]{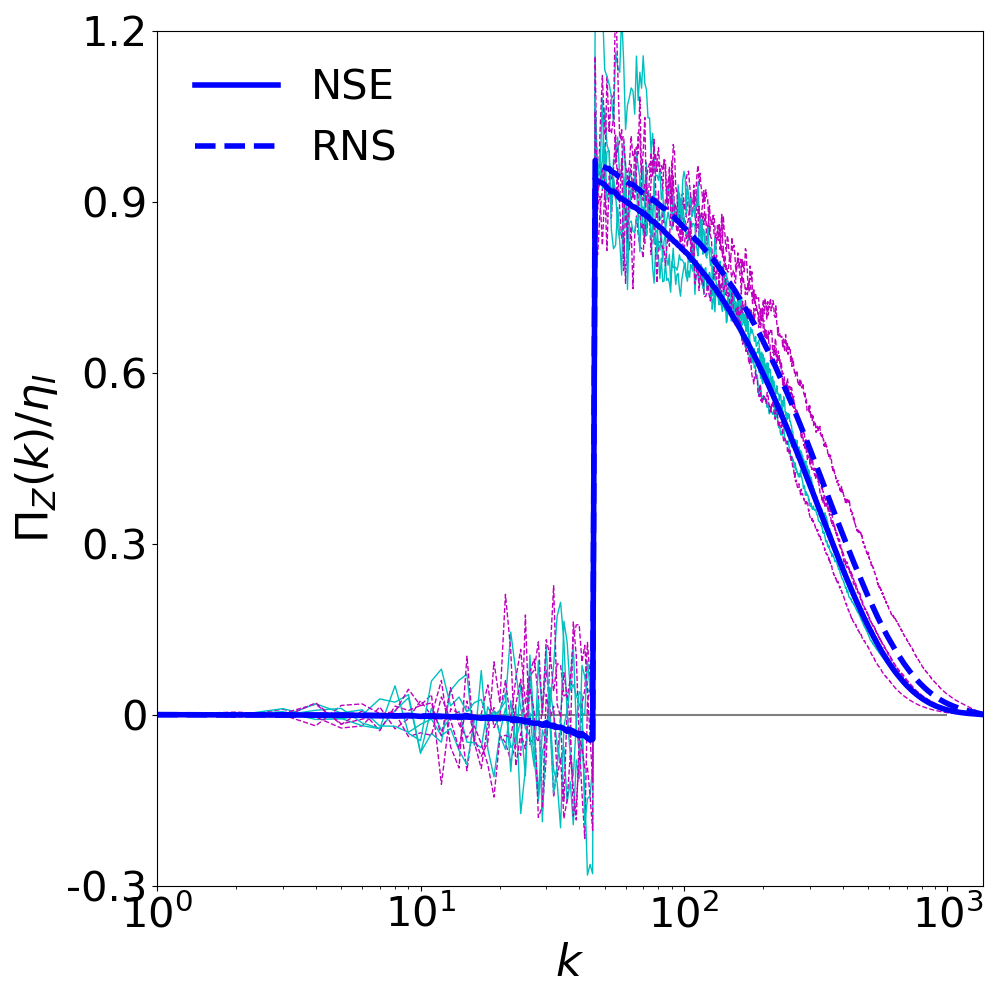}
		\put(-30,190){\Large (c)}
	}
	\caption{Energy spectra and fluxes for the NSE (solid lines) and RNS (dashed lines): (a) energy spectra $E(k)$ exhibits $k^{-5/3}$ and $k^{-4}$ scaling behavior in the inverse and direct cascade regions; (b) energy flux $\Pi_{E}(k)$; (c) enstrophy flux $\Pi_Z(k)$. $\varepsilon_{I}$ and $\eta_I$ are the energy and enstrophy injection rates, respectively. The runs correspond to $\Rey \approx 7.8 \times 10^4$ and $\Rh \approx 2.2 \times 10^4$ with $N^2_c=4096^2$ collocation points}
	\label{fig:specfluxes}
\end{figure*}

Next we characterize the multiscale dynamics of the RNS, and compare it with the NSE, by looking at the spectral quantities such as energy-spectrum and fluxes. 
Figure~\ref{fig:specfluxes} (a) shows the plot of averaged energy spectra obtained for the two systems, on which two inertial ranges are clearly observed. For the inertial range at low wavenumbers, $1 \ll k < k_f$, the spectrum scales as $E(k) \sim k^{-\beta_1}$ with $\beta_1 \simeq 5/3$. The second inertial range with $E(k) \simeq k^{-\beta_2}$ is present over the wave numbers $k_f < k \ll k_{\nu}$ with $\beta_2 \simeq 4$; the latter differs from Kraichnan's prediction of $k^{-3}$ similar to what has been observed for intermediate Reynolds numbers in 2D turbulence \citep{boffetta2010evidence}. $k_{\nu}$ marks the beginning of the wavenumber region where viscous dissipation starts to dominate. Moreover, the slight departure of the exponent $\beta_1$ from $5/3$, seen in the form of build-up of the spectrum at small k-values, can be attributed to the pileup of energy at $k = 1$ which alters the spectral exponent. 

The aforementioned two inertial ranges are associated with the presence of an inverse cascade of energy and a forward (direct) cascade of enstrophy. The plots of energy flux $\Pi_E(k)$ vs $k$ and enstrophy flux $\Pi_Z(k)$ vs $k$, Fig.~\ref{fig:specfluxes} (b) and (c), respectively, confirm this picture for the RNS as well. The energy flux is negative for $k < k_f$ and is almost a constant $\Pi_E(k) \sim \epsilon_I$ in the inertial range at large scales. The enstrophy flux is positive for $k > k_f$ with a tendency to become constant $\Pi_Z(k) \sim \eta_I$ in the inertial range, before starting to rapidly fall at large wavenumbers. Note that the plateau in these fluxes is obtained only in the limit $\Rey \rightarrow \infty, \Rh \rightarrow \infty$. 

\begin{figure*}
	\resizebox{\linewidth}{!}{%
	\includegraphics[scale=0.32]{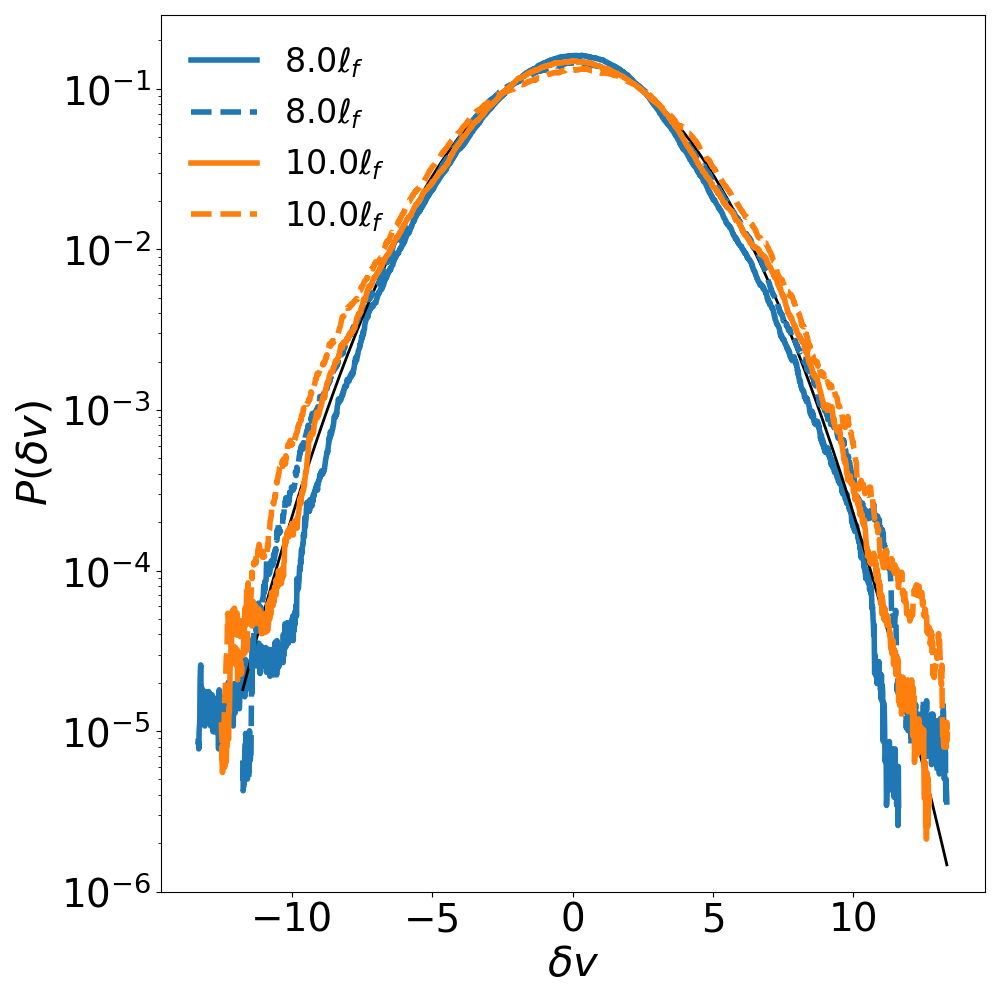}
	\put(-35,190){\huge (a)}
	\includegraphics[scale=0.32]{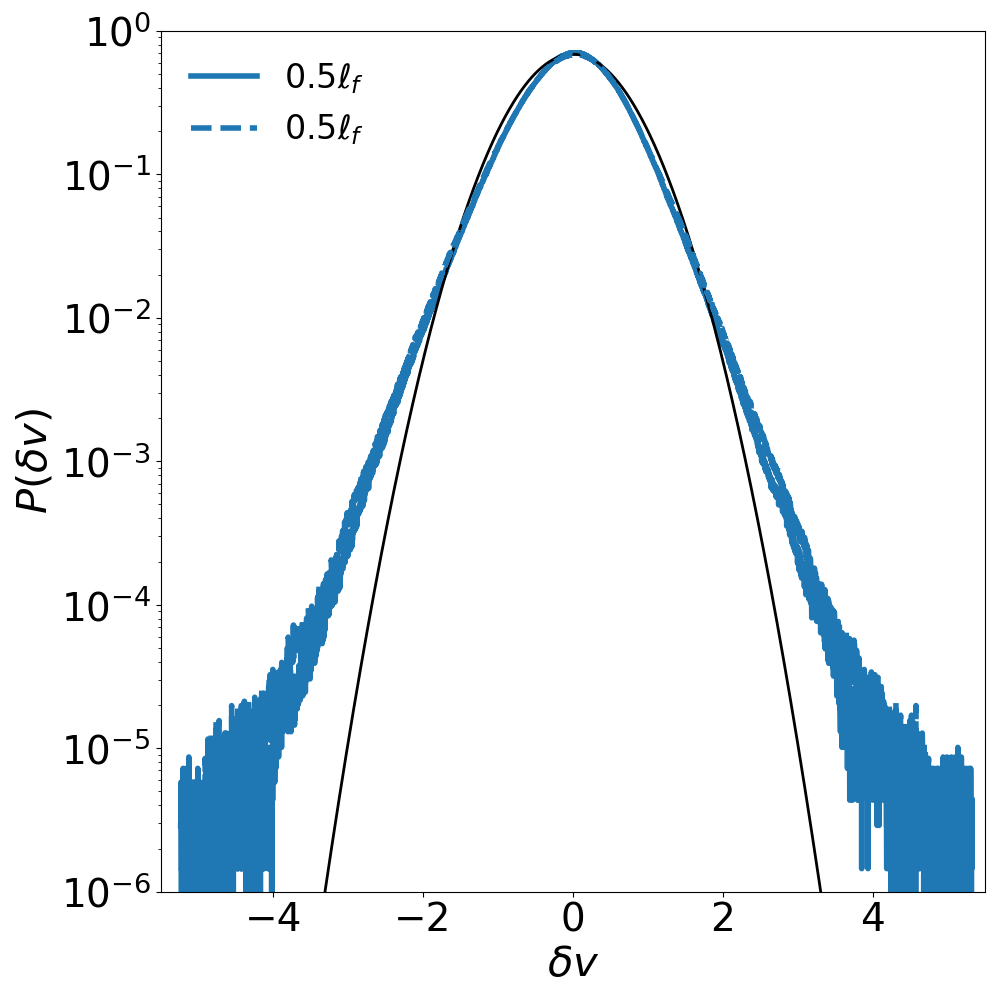}
	\put(-35,190){\huge (b)}
	\includegraphics[scale=0.32]{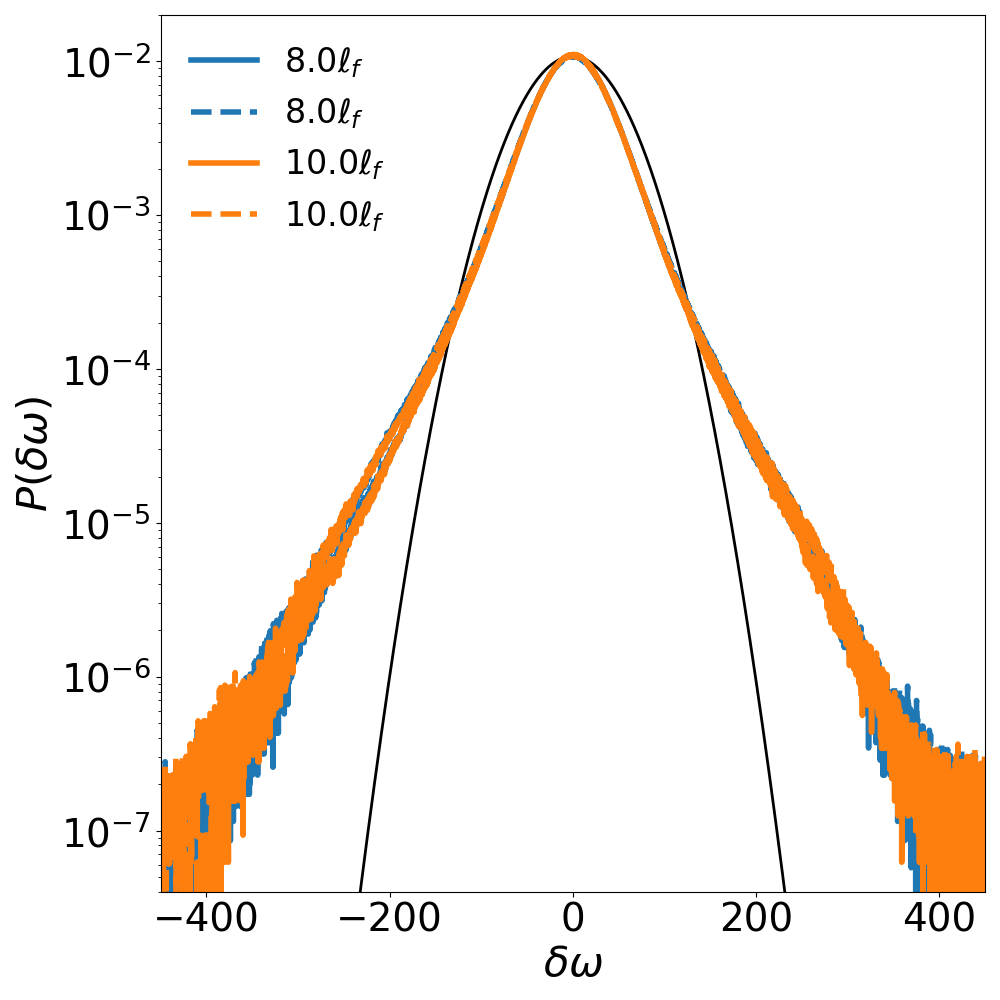}
	\put(-35,190){\huge (c)}	
	\includegraphics[scale=0.32]{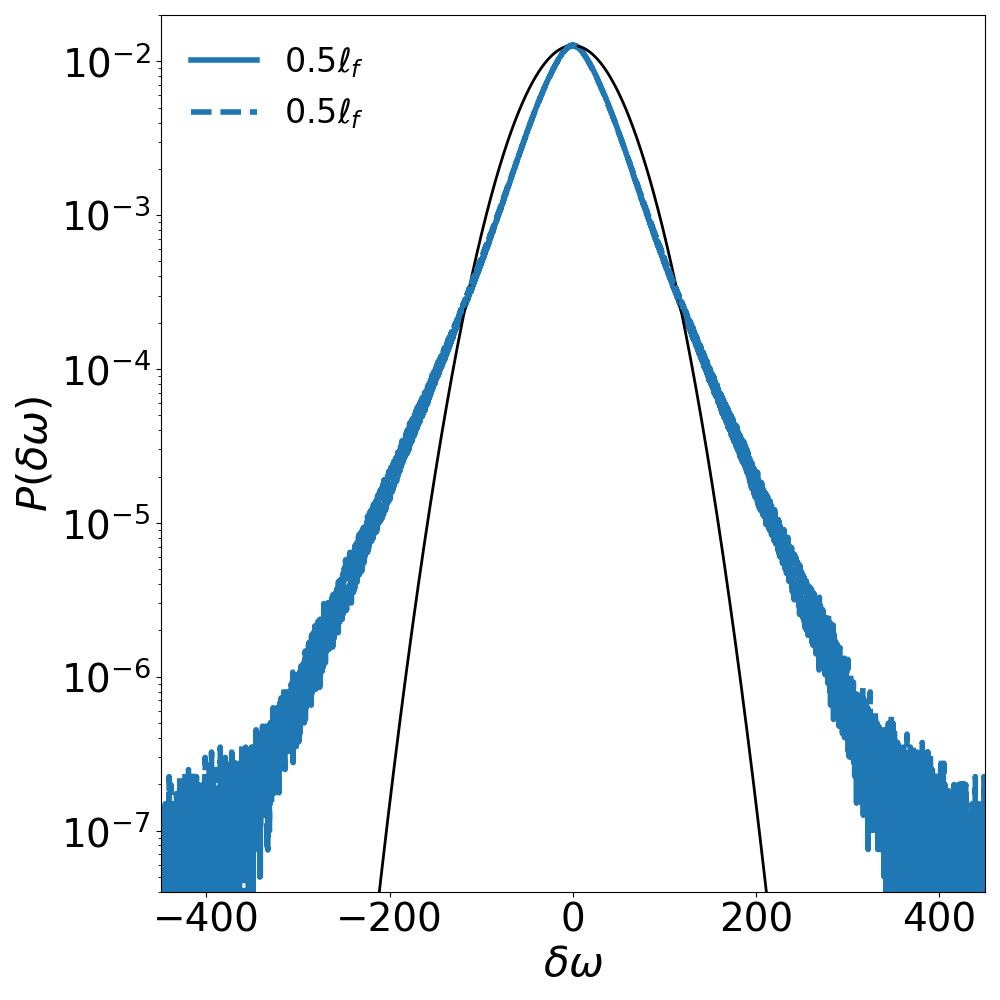}
	\put(-35,190){\huge (d)}	
	}
\caption{Comparison of the probability distribution functions (PDFs) of longitudinal-increments of velocity $\delta v$ and  vorticity $\delta \omega$, of the NSE (solid lines) and RNS (dashed lines), at different length scales in the inverse and direct cascade regions. $\ell_f$ is the forcing length scale; therefore, $\ell > \ell_f$ indicates the inverse cascade region and $\ell < \ell_f$ indicates the direct cascade region. On each of the plots, black solid line represents a Gaussian fit to PDF at the smallest-scale shown for the RNS system. Here $\Rey \approx 7.8 \times 10^4$ and $\Rh \approx 2.2 \times 10^4$ with $N^2_c=4096^2$ collocation points.}
\label{fig:pdfincrem}
\end{figure*}

\begin{figure}
\resizebox{\linewidth}{!}{%		
\includegraphics[scale=0.27]{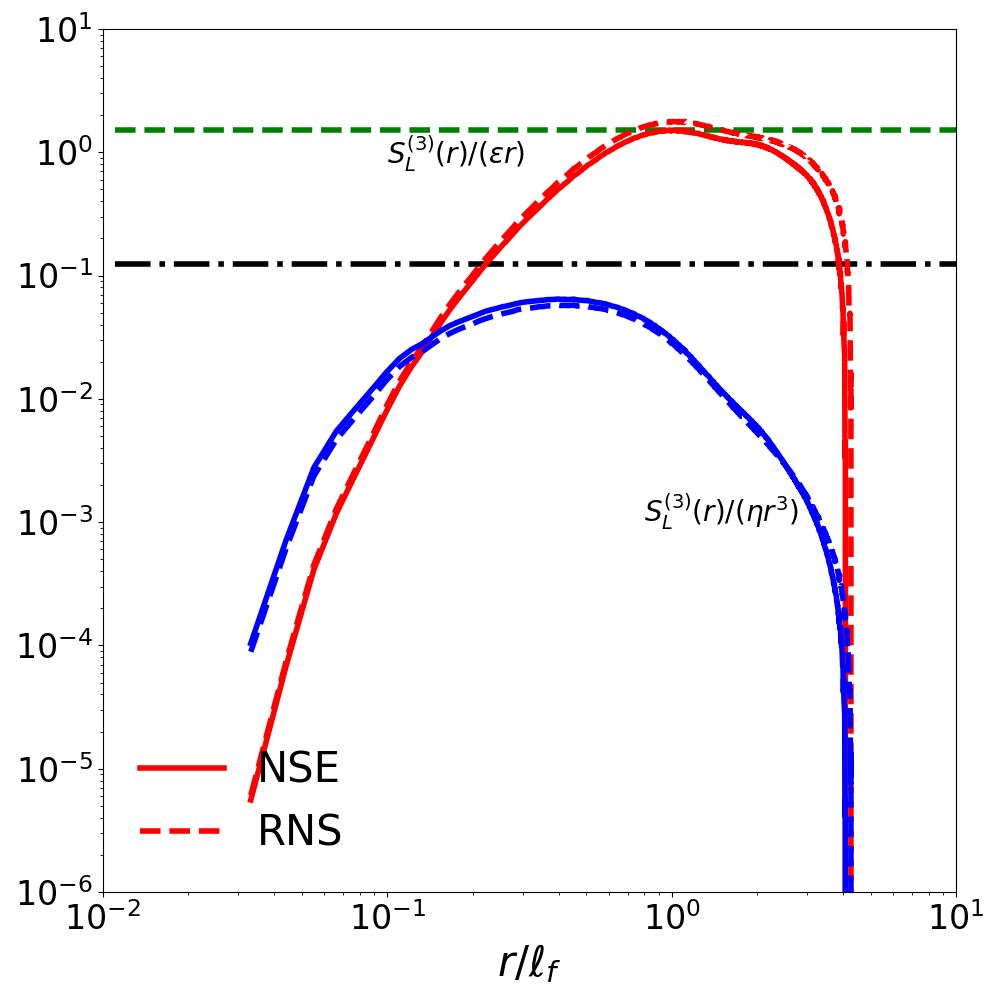}
\put(-98,161){\tiny I}
\put(-48,89){\tiny I}
}
\caption{Plots of compensated third-order longitudinal structure function $S^{(3)}_L$ for the NSE (solid lines) and RNS (dashed lines): Inverse cascade region with plateau at $3/2$ is shown in red color; direct cascade region with plateau $\sim 1/8$ is shown in blue color. $\ell_f$ is the forcing length scale; $\varepsilon_I$ and $\eta_I$ are the energy and enstrophy injection rates, respectively. Green dashed and black dash-dotted lines indicate constant values of $3/2$ and $1/8$, respectively. The runs correspond to $\Rey \approx 7.8 \times 10^4$ and $\Rh \approx 2.2 \times 10^4$ with $N^2_c=4096^2$ collocation points.}
\label{fig:sf}
\end{figure}	

Next we examine the two systems closely using the two point statistics.
The plots of the third-order structure functions for the NSE and RNS are shown in Fig \ref{fig:sf}. They show a similar scaling behavior at both large and small $r$ values. The RNS system exhibits a $S^{(3)}_L \sim r$ and $S^{(3)}_L \sim r^3$ scaling behavior in the inverse and direct cascade regions, respectively. This is in agreement with the well established behavior of the NSE in the dual cascade setting. Moreover, Fig. \ref{fig:sf} clearly shows that the compensated structure function $S^{(3)}_{L}/\left( \epsilon_{I} r \right) \approx 3/2$ both for the RNS and NSE system in the inverse cascade region. Similarly, in the direct cascade region the compensated structure function $S^{(3)}_{L}/\left( \eta_{I}r^3 \right) \approx 1/8$ for both the systems.

We compute the PDFs of the longitudinal velocity and vorticity increments at different length scales for the run with $\Rey \approx 7.8 \times 10^4, \Rh \approx 2.2 \times 10^4$. Figure~\ref{fig:pdfincrem} (a) and (b) show the plots of $\mathcal{P}(\delta v)$ in the inverse and direct cascade regions, respectively. 
These plots, at different length scales greater than $\ell_f$, for the RNS and NSE systems overlap with each other and are very well fitted by a Gaussian distribution, thereby indicating a lack of velocity intermittency. Lack of exact overlap for $\ell=8 \ell_f, 10 \ell_f$ can be attributed to the lack of very good statistical averaging at large length scales. For $\ell < \ell_f$, we observe that the tails of the PDFs deviate from Gaussian distribution. This agreement of the velocity-increment statistics for the two systems also extends to the vorticity increments both in the inverse and direct cascade regions, see Fig.~\ref{fig:pdfincrem} (c) and (d), respectively, where PDFs exhibit exponential tails. We have checked these features at different $\Rey$, while keeping $\Rh$ fixed at a large value of $\sim 2.2 \times 10^4$, and found a very good agreement between the statistics of the two systems, especially in the large $\Rey$ and $\Rh$ limit which can be regarded as the thermodynamic limit with a large number of excited modes (plots not shown for smaller $\Rey$).

\begin{figure*}
\resizebox{\linewidth}{!}{%	
\includegraphics[scale=0.165]{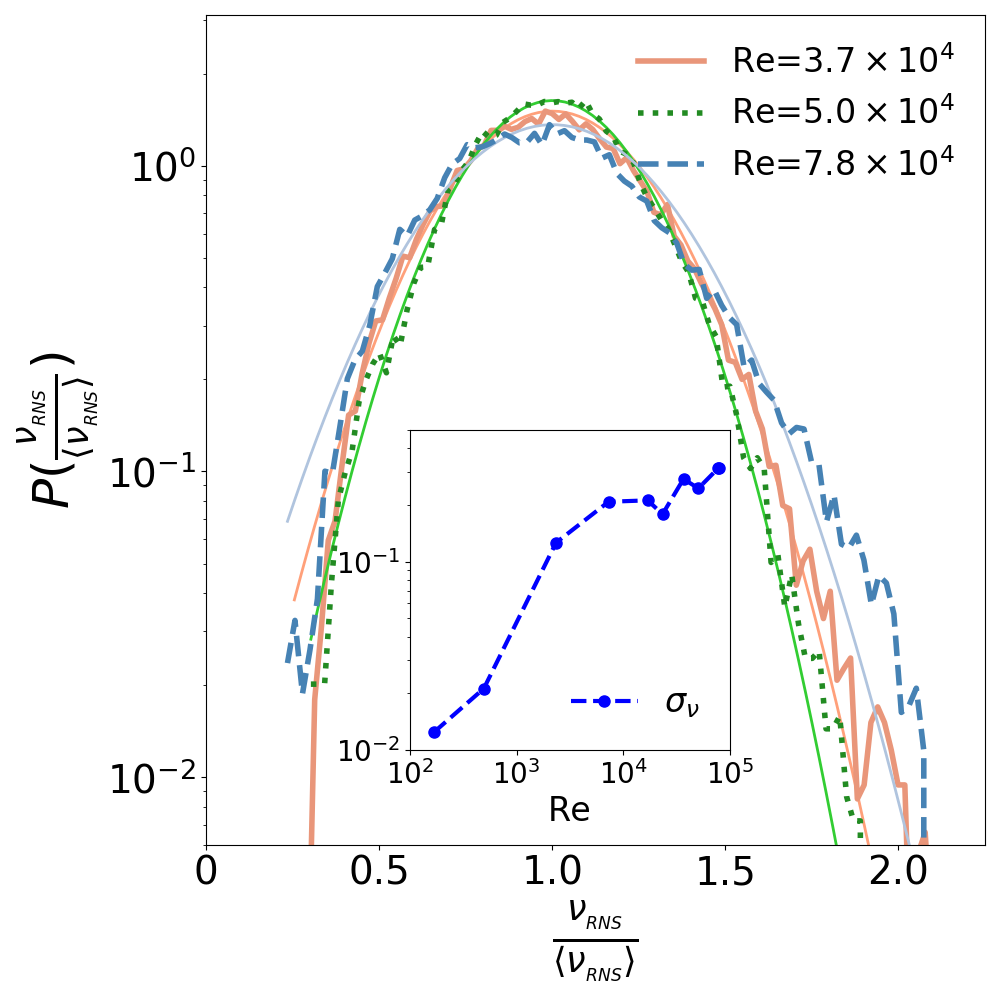}
\put(-90,105){\tiny (a)}
\includegraphics[scale=0.165]{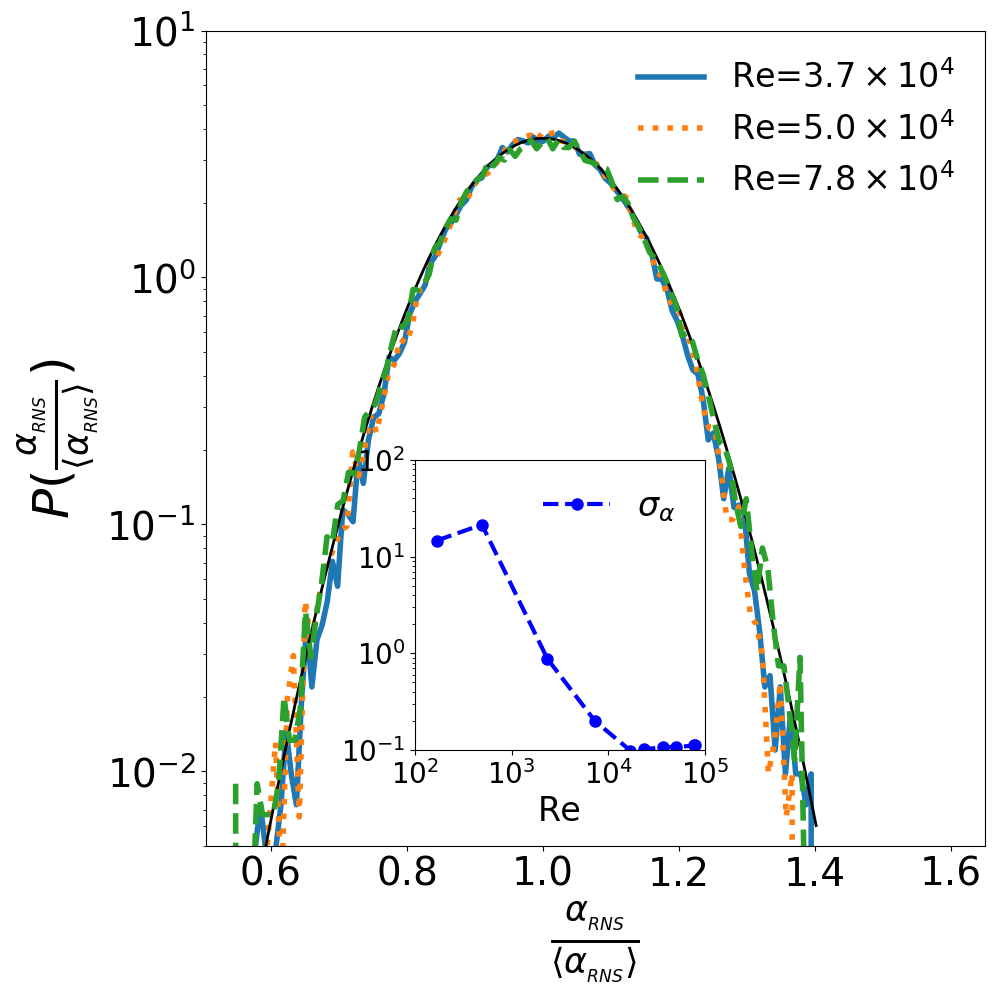}
\put(-90,105){\tiny (b)}
}
\caption{Probability distribution functions of rescaled viscosity $\nu_{_\text{RNS}}/\left\langle \nu_{_\text{RNS}} \right\rangle$ and large-scale dissipation $\alpha_{_\text{RNS}}/\left\langle \alpha_{_\text{RNS}} \right\rangle$ for different $\Rey$, while $\Rh$ is held fixed at $\approxeq 2.2\times 10^4$, for the RNS systems. Insets of these plots show the behaviour of the standard deviation $\sigma_{\nu}$ and $\sigma_{\alpha}$ as a function of $\Rey$. Gaussian fits to PDFs in (a) are indicated by solid of lines of same color, but in lighter shade; in (b) Gaussian fit is indicated by a black solid line.}
\label{fig:pdf_nualpha}
\end{figure*}

Now we focus on the distribution of fluctuations of the dissipation coefficients of the RNS system.
Figure~\ref{fig:pdf_nualpha} shows the PDFs of the rescaled viscosity $\nu_{_\text{RNS}}/\left\langle \nu_{_\text{RNS}} \right\rangle_t$ and rescaled large-scale friction $\alpha_{_\text{RNS}}/\left\langle \alpha_{_\text{RNS}} \right\rangle_t$ for different values of $\Rey$ for the RNS system. Rescaling here is done with respect to the mean of the diffusion coefficients. 
The figures show that both the viscosity and large scale friction have a distribution that is close to Gaussian. It is important to note that the left tail of the distribution of viscosity is always sub-Gaussian. For small values of $\Rey$ the large scale friction has negative events, while at large values of $\Rey$ the large scale friction is always positive over the whole duration of the simulation. 
The viscosity $\nu_{_\text{RNS}}$ on the other hand always remains positive in the parameters that were explored as part of this study. 

Insets of Figs. \ref{fig:pdf_nualpha} show the standard deviation of the rescaled viscosity $\nu_{_\text{RNS}}/\left\langle \nu_{_\text{RNS}} \right\rangle$ and rescaled large-scale friction $\alpha_{_\text{RNS}}/\left\langle \alpha_{_\text{RNS}} \right\rangle$ as a function of $\Rey$.
We see that the standard deviation of the rescaled viscosity is below $1$ even at large $\Rey$ whereas the standard deviation of the rescaled large-scale friction saturates to a value close to $0.1$.

\section{Conclusions}
\label{sec:conclusions}

We have examined the validity of the \textit{equivalence conjecture} for 2D turbulence at moderate to large Re and Rh, while simultaneously imposing two global constraints: conserved total energy and enstrophy. These two quadratic quantities are conserved in the inviscid limit (Euler equation). We find that the RNS system so constructed is able to capture the general behaviour of the NSE system. The global quantities and the spectral diagnostics, one-point statistics, such as energy spectra and fluxes for the two systems agree very well with each other at all Re and Rh. In fact, the modification of the energy spectra at large scales because of the condensate is described very well by the RNS system. 

Fluctuations of the state dependent dissipation coefficients, viscosity and the large scale friction, is well fitted by a Gaussian distribution, except for the observation that the left tail of the viscosity PDF starts to drop suddenly at a large distance from the mean resulting in a slightly asymmetric distribution. This observation could be seen in the light of the fact that we do not observe any negative viscosity event in our simulations. 

Here, our results based on two-point statistics, such as longitudinal velocity structure functions and PDFs of increments of velocity and vorticity, demonstrate that the two systems are statistically similar. Moreover, the agreement between the two systems holds very well for a wide range of $\Rey$, at least for those considered in this study. It remains to be seen whether the formally time-reversible system can capture the dynamics of systems where the form of dissipation is not well defined and whether they can be used to study complex transitions in other turbulent flows \cite{alexakis2018cascades}. 

\textbf{Acknowledgements} Authors acknowledge support from the Institute Scheme for Innovative Research and Development (ISIRD), IIT Kharagpur, Grant Nos. IIT/SRIC/ISIRD/2021-2022/03 and IIT/SRIC/ISIRD/2021-2022/08. V. Shukla   acknowledges  support  from  the  Start-up  Research  Grant No.  SRG/2020/000993 from SCIENCE \& ENGINEERING RESEARCH BOARD (SERB), India.    Authors acknowledge National Supercomputing Mission (NSM) for providing computing resources of ‘PARAM Shakti’ at IIT Kharagpur, which is implemented by C-DAC and supported by the Ministry of Electronics and Information Technology (MeitY) and Department of Science and Technology (DST), Government of India, along with the following NSM Grants DST/NSM/R\&D\_HPC\_Applications/2021/03.21 and DST/NSM/R\&D\_HPC\_Applications/2021/03.11. 

\textbf{Author information} List of authors is arranged alphabetically.

%\bibliographystyle{apsrev4-1}%
%\bibliography{ref}


\end{document}